\documentclass[pdflatex,sn-mathphys-num]{sn-jnl}% Math and Physical Sciences Numbered Reference Style
\usepackage{graphicx}%
\usepackage{multirow}%
\usepackage{amsmath,amssymb,amsfonts}%
\usepackage{amsthm}%
\usepackage{mathrsfs}%
\usepackage[title]{appendix}%
\usepackage{xcolor}%
\usepackage{textcomp}%
\usepackage{manyfoot}%
\usepackage{booktabs}%
\usepackage{algorithm}%
\usepackage{algorithmicx}%
\usepackage{algpseudocode}%
\usepackage{listings}%
\usepackage{lmodern}
\theoremstyle{thmstyleone}%
\theoremstyle{thmstyletwo}%

\theoremstyle{thmstylethree}%

\begin{document}

\title{Bistability to Quad-stability: Emergence of Hybrid Phenotypes \& Enhanced Spatio-temporal Plasticity in Presence of Host-Circuit Coupling}
%\title{Host-Circuit Coupling in Toggle Switch: Emergence of Hybrid States, Quad-stability \& Spatio-temporal Patterns}


\author[1]{\fnm{Ranu} \sur{Kundu}}%\email{iauthor@gmail.com}

\author[2]{\fnm{Priya} \sur{Chakraborty}}%\email{iiauthor@gmail.com}

\author[1]{\fnm{Sohini} \sur{Guin}}%\email{iiiauthor@gmail.com}
\author[1]{\fnm{Shyam Sundar} \sur{Poriah}}%\email{iiiauthor@gmail.com}

\author[1]{\fnm{Sayantari} \sur{Ghosh}}
\email{sghosh.phy@nitdgp.ac.in}
\affil[1]{\orgdiv{Department of Physics}, \orgname{National Institute of Technology Durgapur},\city{Durgapur} \postcode{713209}, \country{India}}

\affil[2]{\orgdiv{Department of Physics}, \orgname{Indian Institute of Technology Bombay}, \city{Mumbai} \postcode{400076},\state{Maharashtra}, \country{India}}
%\affil[2]{\orgdiv{Department}, \orgname{Organization}, \orgaddress{\street{Street}, \city{City}, \postcode{10587}, \state{State}, \country{Country}}}

%\affil[3]{\orgdiv{Department}, \orgname{Organization}, \orgaddress{\street{Street}, \city{City}, \postcode{610101}, \state{State}, \country{Country}}}


\abstract{In the context of multistability-driven diseases, like cancer, spatio-temporal plasticity plays a significant role to achieve a spectrum of phenotypic variations. The interplay between gene regulatory networks and environmental factors, such as resource competition and spatial diffusion, plays a crucial role in determining cellular behaviour and phenotypic heterogeneity. Though reaction–diffusion frameworks have been widely applied in developmental biology, less attention has been paid to the simultaneous effects of resource competition and growth feedback on spatial organization. In this paper, we observed that a bistable genetic circuit under high resource competition due to growth feedback gives rise to multiple emergent phenotypes, as observed in cancer systems. Furthermore, we observed how spatial diffusion coupled with intrinsic nonlinearity can drive the emergence of distinct spatial dynamics over time. The observed spatiotemporal plasticity can also be driven by the comparative stability of the fixed points, diffusivity, and asymmetry of diffusion. Our findings highlight that growth-induced resource competition combined with diffusion can provide deeper insights into metastasis and cancer progression.}


\maketitle

\section{Introduction}\label{sec1}

In gene regulatory circuits, resource competition plays a crucial role in understanding the dynamics of gene expression within the host cell \cite{frei2020characterization, gyorgy2018sharing, cookson2011queueing, di2023resource}. Asymmetric utilization of significant cellular resources  (such as ribosomes, RNA polymerase, proteases, etc.) can disrupt the cellular economy and affect genetic functions, impacting overall cellular homeostasis  \cite{qian2017resource,gyorgy2015isocost}.  Many researchers demonstrated how growth feedback with circuit interaction results in unexpected outcomes  \cite{tan2009emergent,zhang2020topology,melendez2021emergent, ghosh2012emergent}. 
%\textcolor{red}{Therefore, it has been established that resource competition not only affects cellular function but also has significant implications for the behaviour of regulatory circuits } Apart from synthetic biology, resource competition plays a prominent role in diseases like cancer, diabetes, COPD etc. \cite{trainor2014ribosome, lomas2016does, thomson2016endogenous}.\\
If we consider the case of cancer specifically, then it should be noted that the uncontrolled growth of cancer cells demands increasing amounts of cellular resources, oxygen, nutrients, and growth factors \cite{fadaka2017biology, papalazarou2021supply}. Thus, the metabolic burden by growth exerts a diverse impact on gene circuits  \cite{klumpp2014bacterial,tan2009emergent}. In tumor systems, competition can occur for cellular resources as well as for nutrients in the early stages \cite{hockel2001tumor,pianka1981competition}, as diffusion allows only approximately 1 mm depth to reach from surrounding blood capillaries \cite{boucher1992microvascular}. Thus, the primary site of cancer/tumor cells actually consists of some densely packed cells competing for scarce resources \cite{garcia2019starvation}. Metastasis, the migration of cells from their primary location to a secondary location \cite{chambers2002dissemination, gupta2006cancer}, is also regulated by this resource scarcity in the primary location \cite{joyce2009microenvironmental, kim2023cell}. This motility does not always ensure cell fitness outside the primary location, as the majority of cells die rather than initiate a secondary tumor, but this migration effectively ensures the fitness of the primary tumor with a chance of secondary formation as well \cite{taylor2017resource}. Further insights into the role of nutrient limitation in cancer evolvement have been established in \cite{aktipis2012dispersal}, and resource heterogeneity in cancer systems have been pointed out \cite{chen2011solving}. \\
Like many other diseases \cite{golos2015multistability,ghosh2011phenotypic}, cancer is also driven by inherent nonlinearity and multistable dynamics, which causes EMT (\textcolor{black}{Epithelial-to-Mesenchymal Transition})-driven cellular plasticity. \textcolor{black}{Cellular plasticity refers to the ability of cells to reversibly transition between distinct phenotypic states in response to internal regulatory fluctuations or external environmental cues, without permanent genetic alterations. EMT is one of the most prominent manifestations of such plasticity, wherein stationary epithelial cells acquire mesenchymal characteristics.} Classical bistable switch, the simplest depiction of multistable dynamics in genetic systems, is essentially a type of gene regulatory network consisting of two mutually inhibiting transcription factors (say U and V). This circuit is associated with binary cell fate decision i.e. existence of U-high/V-low and U-low/V-high state \cite{gardner2000construction,macia2009cellular,andrecut2011general}. \textcolor{black}{%representing extreme or terminal phenotypes: Epithelial (strong
%cell-cell adhesion) and Mesenchymal (to migrate). 
For example, in cancer, evidence shows that epithelial–mesenchymal transition (EMT) is regulated by mutually inhibitory miRNA-transcription factor (TF) circuits, such as the miR-200/ZEB and miR-34/SNAIL networks \cite{tian2013coupled, jolly2015implications}. In these systems, high miRNA expression promotes epithelial traits, whereas high transcription-factor expression represses E-cadherin (a hallmark of the epithelial phenotype) and promotes mesenchymal traits. Recently, along with two extreme phenotypes, the existence of an intermediate hybrid E/M state (cluster cell migration) has been established \cite{sinha2020emerging}. In the hybrid state, cancer cells do not lose their epithelial property completely but gain the migratory trait \cite{christiansen2006reassessing, klymkowsky2009epithelial}. This dual nature significantly enhances their malignant potential, enhancing their capacity for metastasis, and driving more aggressive cancer progression} \cite{pastushenko2018identification,jolly2015implications,gupta2019phenotypic}. %It has been shown that if one or both TFs of toggle switch strongly auto-activate itself as well as repress the other, the circuit may give rise to these three distinct phenotypes\cite{lu2013tristability,jolly2015implications}.  
Recently it has also been explored that there are multiple hybrid states depending on the ratio of epithelial and mesenchymal traits the cells can exhibit \cite{pastushenko2019emt,kroger2019acquisition}. The origin of these hybrid states have been modeled using the existence of strong autoregulations \cite{jolly2015implications}, multi-level boolean models \cite{hari2024multi}, poor vascularization \cite{tretyakova2022tumor}, hypoxic stress in the tumor microenvironment \cite{saxena2020hypoxia} etc. \textcolor{black}{These observations highlight that EMT-associated phenotypic plasticity is not a static or purely binary process, but rather reflects the ability of cancer cells to reversibly transition among epithelial, mesenchymal, and intermediate hybrid states \cite{jolly2015implications}. Such plasticity enables cells to dynamically adapt to fluctuating microenvironmental cues \cite{friedl2010plasticity}. Importantly, this adaptive behavior manifests across both space and time within tumors: spatially due to gradients of nutrients, oxygen, and growth factors, and temporally due to dynamic reprogramming of gene regulatory networks and cell–cell interactions. This spatio-temporal plasticity is a key contributor to tumor heterogeneity, metastatic progression, and treatment resistance \cite{pinto2013breast,nieto2013epithelial,corallino2015epithelial}.}\\
In this paper, our analysis suggests that a simple bistable genetic switch, operating under limited intracellular gene expression resources, can give rise to multiple intermediate phenotypes similar to those observed in the cancer system. Establishing the connection of resource limitation to the growth of the cell \textcolor{black}{(which are fundamental hallmarks of cancer, as discussed above)}, we consider that while natural degradation rates are mostly linear, in a resource-limited environment, the degradation can be a complex nonlinear function. Inspired by mathematical
models for incorporating the effect of this nonlinear decay through growth modulation \cite{ghosh2011phenotypic, ghosh2012emergent,tan2009emergent}, we revisit the dynamics of a bistable genetic circuit in a resource-limited condition. To extend the analysis, we also incorporated diffusion along with this intrinsic nonlinearity and observed how these multiple steady states respond, coupled with diffusion, leading to spatial heterogeneity, thereby forming distinct spatial organization with time. This indicates that resource limitation combined with spatial factors gives rise complex tissue level patterning. %While following Turing's seminal studies \cite{turing1990chemical}, the emergence of complex patterns in developing organisms through various mechanisms and chemical reactions  \cite{ vittadello2021turing,kondo2010reaction,painter1999stripe} has been well-studied, pattern formation in bistable systems, like toggle switches remain largely unexplored. 
Tissue pattern formation plays a crucial role in the development and maintenance of healthy tissues and organs in multicellular organisms. Instead of appearing and growing in a well-organized manner like healthy tissues, cancerous tissue often shows irregular patterns, disorganization and disruption in their structure \cite{marongiu2012cancer}. Researchers have also proposed that this altered morphology is fundamental to the formation of the tumor microenvironment, which is a crucial factor in metastasis. Therefore, in this work, we attempt to understand the mechanism of occurrence of multistable hybrid states and associated spatiotemporal plasticity caused by a simple bistable genetic switch in resource limitation conditions, mediated by nonlinear growth feedbacks.\\

\section{Model Formulation}%\label{section 1}
\begin{figure} [ht]
    \centering
    \includegraphics [width=1\linewidth]{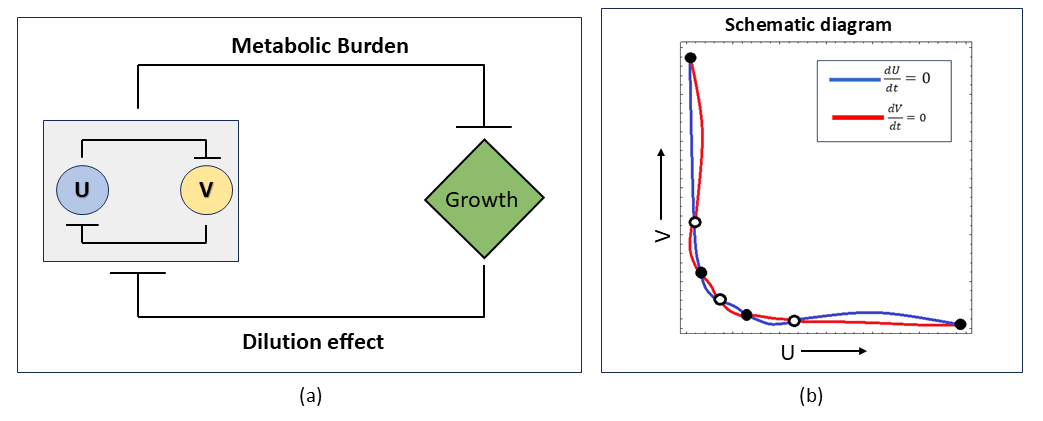}
    \caption{(a) Schematic representation of the mutual inhibitory loop between the bistable genetic switch and cellular growth. Proteins $U$ and $V$ mutually repress each other, forming a bistable genetic circuit, which is further coupled to growth feedback through metabolic burden and dilution effects, thereby establishing a mutual inhibitory loop. (b) Schematic nullcline diagram for $U$ (blue) and $V$ (red). Their intersections correspond to the system’s steady states, yielding a total of seven fixed points. Filled circles denote stable fixed points, while hollow circles indicate unstable fixed points.}
    \label{model}
\end{figure}

Let us consider a mutual inhibitory loop between two proteins $U$ and $V$, where each represses the synthesis of the other. A lower concentration of $U$ and a higher concentration of $V$ lead cells to adopt epithelial characteristics, and a higher concentration of $U$ and a lower concentration of $V$ drive cells toward a mesenchymal phenotype. Here, we consider the cooperativity  ($n$) of both $U$ and $V$ to be 2, indicating that repression is mediated by dimer formation of the respective proteins. Basal synthesis rate of the protein $U$ and $V$ are represented by $K_{0u}$ and $K_{0v}$ respectively. $K_{1u}$ and $K_{1v}$ are the maximum inducible expression rate of protein $U$ and $V$. The maximum dilution rate caused by cell growth has been considered as $K_{GR}$, which indicates the growth feedback rate. Now, following the proposal of \cite{tan2009emergent,chakraborty2023spatio}, to incorporate the interaction between the host cell and the genetic circuit, we take into account that the production of proteins $U$ and $V$ requires essential cellular resources (like RNA polymerases, ribosomes etc.). Overexpression of these proteins draws resources vital for other cellular functions,  imposing a metabolic burden on cell growth. Conversely, as the host cell grows, the concentration of expressed gene products decreases due to the dilution effect, thereby reducing its overall impact. These two mutual processes negatively affect the other, resulting in a double negative feedback in the host-circuit system, as shown in Fig. \ref{model}(a). The last term of the differential equations [Eq. (\ref{eqn 1}) and (\ref{eqn 2})] given below represents the nonlinear reduction in the dilution rate due to this host-circuit coupling. The natural 
degradation rate has been considered as $d_{0u}$ and $d_{0v}$ for $U$ and $V$, respectively. The above considerations are mathematically represented by these equations:\\
\begin{equation}
    \frac{dU}{dt}= K_{0u} + \frac{K_{1u}}{K_u^n + V^n} - d_{0u}\;U - \frac{K_{GR}}{K_c (U+V)^m + 1}\,U
    \label{eqn 1}
\end{equation}
\begin{equation}
    \frac{dV}{dt}= K_{0v} + \frac{K_{1v}}{K_v^n + U^n} - d_{0v}\;V - \frac{K_{GR}}{K_c (U+V)^m + 1}\,V
    \label{eqn 2}
\end{equation}
In the above equation we assumed $K_{c}\,= 1/J^m$, where $J$ reflects the expression capacity for the genes $U$ and $V$, while the Hill coefficient $m$ represents the sensitivity of the metabolic burden induced by their expression. We will explore this above-mentioned model for steady states in the next section.\\
To extend the model and examine the evolution of spatio-temporal dynamics in the presence of diffusion, we have next considered a two-dimensional thin (mono-layer) sheet of cells where the motif shown in Fig. \ref{model} (a) is present in each cell. The proteins can diffuse through the cell boundaries with equal rates along $X$ and $Y$ axis  (isotropic diffusion) and the diffusion coefficient is $D_U$ for protein $U$ and $D_V$ for protein $V$. The corresponding equations to be studied further are:
\begin{equation}
    \frac{\partial U (x,y,t)}{\partial t}= K_{0u} + \frac{K_{1u}}{K_u^n + V^n} - d_{0u}\;U - \frac{K_{GR}}{K_c (U+V)^m + 1}U + D_U\;\nabla^2U
    \label{eqn 3}
\end{equation}
\begin{equation}
    \frac{\partial V (x,y,t)}{\partial t}= K_{0v} + \frac{K_{1v}}{K_v^n + U^n} - d_{0v}\;V - \frac{K_{GR}}{K_c (U+V)^m + 1}V + D_V\;\nabla^2V
    \label{eqn 4}
\end{equation}
where 
    $\nabla^2 =\frac{\partial^2}{\partial x^2} +\frac{\partial^2}{\partial y^2}$  is the Laplacian Operator.\\
\\

\section{Results}%\label{section 2}
\subsection{Bifurcation \& Emergence of Multiple Hybrid States:}
\textcolor{black}{We begin by exploring the dynamical behavior of the system. Equating the time rate of change of Eq. (\ref{eqn 1}) and (\ref{eqn 2}) to zero, we get the steady states which correspond to the real solutions of the system. Linear stability of each steady state is determined by evaluating the eigenvalues of the Jacobian matrix at the corresponding fixed points. } We tuned the circuit parameters to visualize the bifurcation behaviour of the protein $U$ (and $V$, not shown here) by varying a control parameter $K_{1u}$ and $K_{GR}$, which represent the synthesis rate of protein $U$ and growth rate of the cell, respectively. 
%To explore the dynamics of the system, we investigate the nullclines corresponding to  Eq. (1) and  (2), and note that the system can give rise to quad-stability, i.e., four distinct stable states, separated by three unstable states \textcolor{red}{[Fig.\ref{model} (b)].} However, in the absence of the resource competition ($K_{GR}=0$), the system gives rise to the conventional bistable genetic switch model, with two stable states.\\

\begin{figure}
    \centering
    \includegraphics [width=1\linewidth]{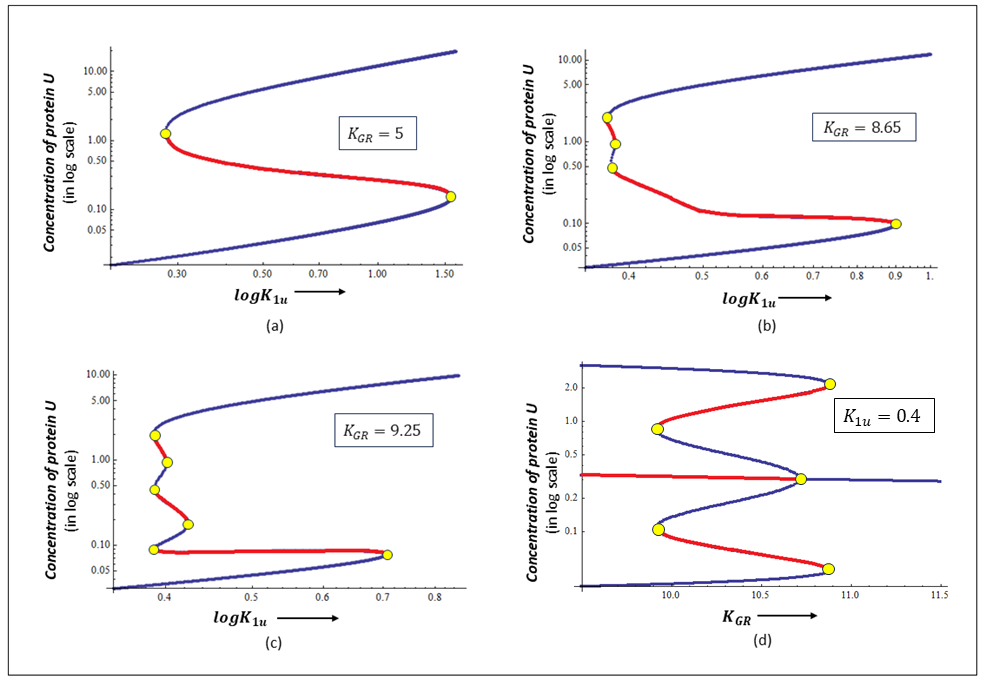}
\caption{(a)-(c) Bifurcation diagram of concentration of protein $U$ (in log scale) with variation of $K_{1u}$ (in log scale) for different growth rates: (a) $K_{GR}=5$  (b) $K_{GR}=8.65$  (c) $K_{GR}=9.25$; Corresponding parameters are: $K_{0u}=K_{0v}=0.005,\; K_{1v}=0.4,\; K_u=K_v=0.09,\; d_{0u}=d_{0v}=0.9,\; K_c=1.6,\; n=2,\; m=2.$   (d) Bifurcation diagram of concentration of protein $U$ (in log scale) with variation of $K_{GR}$ for a fixed protein synthesis rate ($K_{1u}\,=0.4$).
 Other parameters are: $K_{0u}=K_{0v}=0.005,\; K_{1v}=0.4,\; K_u=K_v=0.09,\; d_{0u}=d_{0v}=0.9,\; K_c=1.8,\; n=2,\; m=2.$ Blue lines and red lines indicate the stable and unstable states, respectively, and yellow circles represent the bifurcation points. }   
    \label{bif_dig}
\end{figure}

\begin{itemize}
    \item In the low-growth regime ($K_{GR}\,=5$), the metabolic burden—and hence resource competition—is minimal. Under these conditions, the dynamical system undergoes two saddle-node bifurcations, giving rise to two distinct steady-state concentrations of protein $U$: one low and one high, as the synthesis rate is varied (Fig. \ref{bif_dig} (a)). As a result, the system exhibits bistable behaviour reflecting the binary phenotypic states commonly observed in cancer cells that acquire either epithelial or mesenchymal phenotypes. 
    \item At a moderate growth rate ($K_{GR}\,=8.65$) metabolic burden enhances the resource competition within the system. This results in the emergence of a tristable regime where the concentration of protein $U$ can exist at three distinct levels: low, intermediate, and high [Fig. \ref{bif_dig}(b)]. The intermediate state represents the hybrid state, which is more metastatic as the cancer cells show their epithelial and mesenchymal characteristics simultaneously.
    \item With further increase of growth rate ($K_{GR}\,=9.25$) the resource competition is also raised, enabling cancer cells to bifurcate themselves into four distinct phenotypes: epithelial ($U$ low), mesenchymal ($U$ high), hybrid I (more epithelial-less mesenchymal) and hybrid II states (more mesenchymal-less epithelial) [Fig. \ref{bif_dig} (c)]. These multiple hybrid states are associated with increased metastatic potential that allows the cancer cells to spread more rapidly.
  
\end{itemize}
These bifurcation diagrams [Fig. \ref{bif_dig}(a)-(c)] show how a simple bistable cancer system gives rise to multiple hybrid states in a resource-limited condition. 

\subsection{Diffusion-mediated Spatio-Temporal Dynamics \& Phenotypic Transitions:}
Since Turing’s seminal studies \cite{turing1990chemical}, the emergence of complex spatial organization in developing organisms has been extensively explored through various mechanisms and reaction–diffusion models \cite{vittadello2021turing,kondo2010reaction,painter1999stripe}. However, pattern formation in multistable dynamical systems remains largely unexplored. To address this gap, we investigate how diffusion coupled with nonlinear dynamics drives spatio-temporal organization and phenotypic transitions in our system.
In this section, we study the spatial dynamics of the system by introducing a two-dimensional 200$\times$200 square lattice of cells, where the position of each cell is  ($x_i$, $y_i$) with $x_i$ $\in$  (1, 200) in the x-direction and $y_i$ $\in$  (1, 200) in the y-direction. Under no-flux boundary conditions, we simulated the coupled PDE's (Eq. \ref{eqn 3}, \ref{eqn 4})  using \textcolor{black}{forward difference method} with the mathematical form of the initial conditions given below. 
\begin{equation*}
    U_{initial}(x_i,y_i,0) = \epsilon_1\;\xi (0,1)\;\;\;\; V_{initial}  (x_i,y_i,0) = \epsilon_2\;\xi (0,1)
\end{equation*}
Here, $\xi (0,1)$ represents a random uniform distribution with a scaling factor $\epsilon_1$ for protein $U$ ($\epsilon_2$ for protein $V$). \\

\begin{figure}[ht] 
    \centering
    \includegraphics [width=1\linewidth]{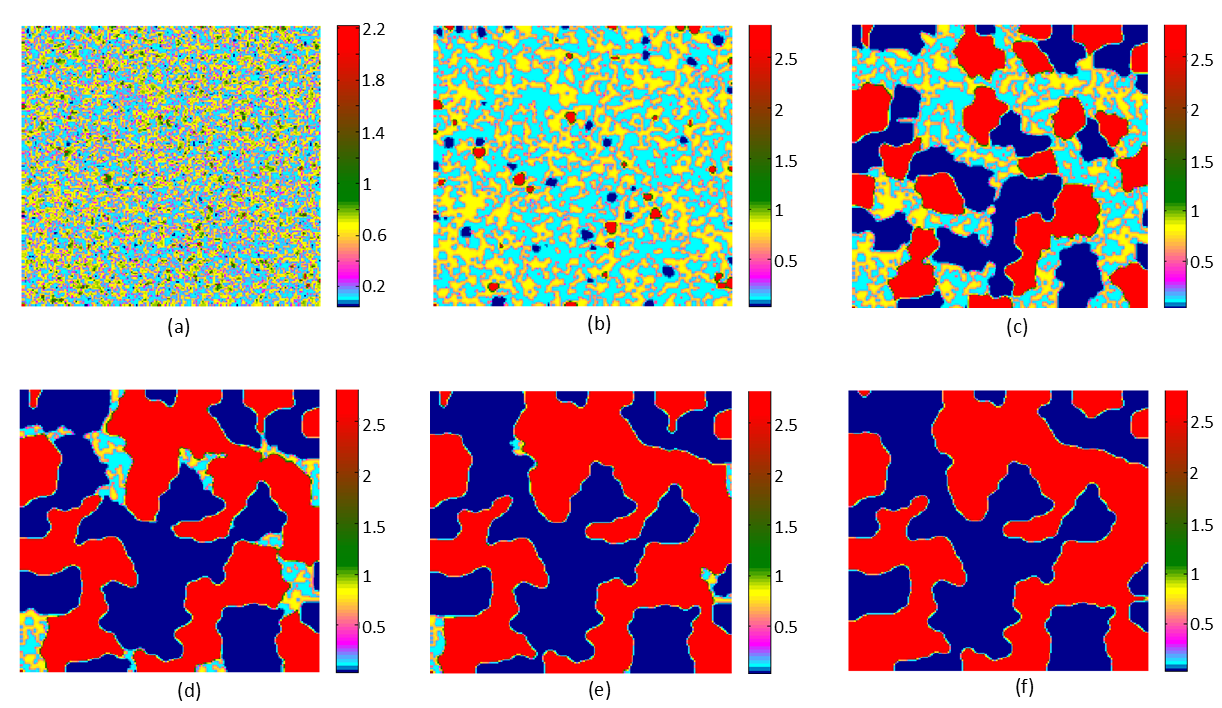}
    \caption{Spatio-temporal pattern formation  (in the presence of diffusion) for protein $U$. Corresponding parameters are: $K_{0u}=K_{0v}=0.005, K_{1u}=K_{1v}=0.4, K_u=K_v=0.09, d_{0u}=d_{0v}=0.9, K_c=1.6,K_{GR}=9.25, n=2, m=2,\; \epsilon_1=\epsilon_2=3$ Diffusion coefficients are $D_U$=$D_V=0.04$. Snapshots are taken after time: (a)$10$  (b)$100$  (c)$500$  (d)$800$  (e)$1000$  (f)$2000$. The colors of associated colorbars indicate different concentrations of protein U. }
    \label{pattern}
\end{figure}
\begin{figure}[ht]
    \centering
    \includegraphics[width=1\linewidth]{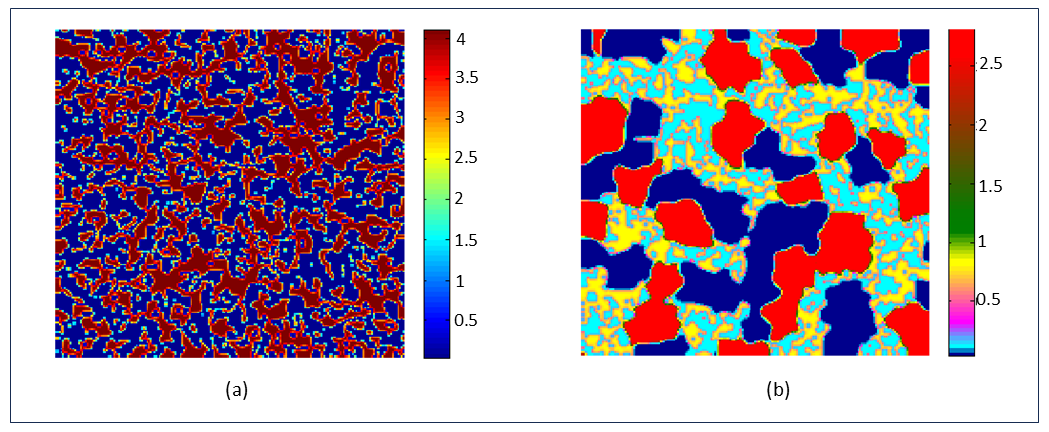}
    \caption{Bistability vs. Quad-stability in spatio-temporal dynamics for protein $U$. (a) The value of $K_{GR}\,=5$ results in bistability as observed in the classical genetic switch. (b) Increasing $K_{GR}\,=9.25$ leads to multistability in the system. Snapshots are taken after time $500$; The parameters are same as Fig. \ref{pattern} .}
    \label{bistable}
\end{figure}

Now, we examine the influence of diffusion coupled with intrinsic nonlinearity on steady-state transitions and spatial dynamics over time. %\textcolor{red}{The selected parameters are: $K_{0u}=K_{0v}=0.005, K_{1u}=K_{1v}=0.4, K_u=K_v=0.09, d_{0u}=d_{0v}=0.9, K_c=1.6,K_{GR}=9.25, n=2, m=2,\; \epsilon_1=\epsilon_2=3$ while the diffusion coefficients are $D_U$=$D_V=0.04$.}
We choose the system parameters from the quadrastable regime of Fig. \ref{bif_dig}(c) in the multicellular set-up of diffusible protein molecules. At the very beginning of the simulation, the system is more reaction-driven [Fig. \ref{pattern}(a)]; over time, two intermediate states become prominent, leading to a shift in the system's behaviour towards bimodality [Fig. \ref{pattern}(b)]. This binary response arises due to the coexistence of hybrid I and hybrid II states together.  As time progresses, two additional extreme states start to emerge, exhibiting the quad-stable dynamics of the system [Fig. \ref{pattern}(c)]. In this phase, four distinct phenotypes are simultaneously present in the system: hybrid I and hybrid II  (two intermediate states) along with two newly formed extreme states, epithelial and mesenchymal. With further progression of time, the epithelial and mesenchymal states undergo continued diffusion, which causes the hybrid states to gradually disappear from the system [Fig. \ref{pattern}(d)- (e)]. Consequently, the system returns back to a bistability, characterized by purely epithelial and mesenchymal cells [Fig. \ref{pattern}(f)], depicting enhanced stability of E or M states.\\
We further investigate and compare the spatiotemporal dynamics of the system in a high and low resource limitation regime side by side. We precisely highlight the absence of hybrid states in Fig. \ref{bistable}(a) (in a low resource limitation regime) at $t=500$ when compared to the spatiotemporal dynamics of Fig. \ref{bistable}(b) (with high resource limitation). Given that the strongest stemness, gene expression occurs at the commencement of the hybrid state, we see that resource limitation plays a role in the development of these hybrid states. 
\\

\begin{figure}
    \centering
    \includegraphics[width=1.0\linewidth]{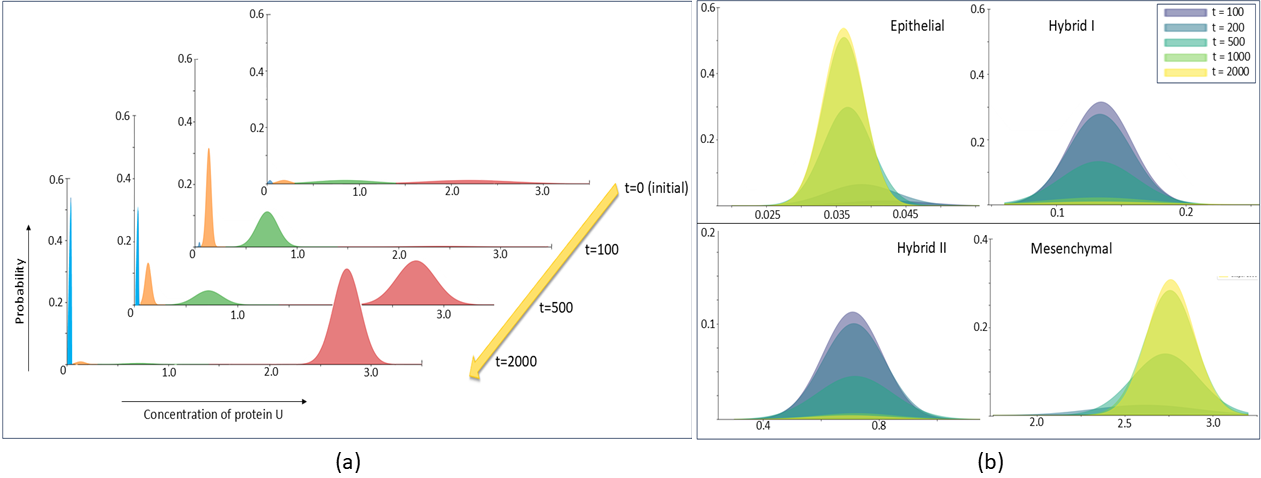}
    \caption{Gaussian stacked distributions illustrating the temporal evolution of protein $U$ concentration. (a) Probability distributions of protein $U$ at different time points ($t= 0,\, 100,\, 500,\, 2000$). The x-axis and y-axis represent the concentration of protein $U$ and the probability of cells at that specific concentration, respectively. (b) Temporal evolution of Gaussian distributions corresponding to the four phenotypic states: epithelial, hybrid-I, hybrid-II, and mesenchymal. Parameter values are taken as: $K_{0u}=K_{0v}=0.005,\; K_{1u}=K_{1v}=0.4,\; K_u=K_v=0.09,\; d_{0u}=d_{0v}=0.9,\; K_c=1.6,\;K_{GR}=9.25,\; n=m=2,\; D_U=D_V=0.04$.}
    \label{stacked plot}
\end{figure}

To quantify the temporal evolution in terms of the spatial distribution, we have analyzed the probability distribution of the protein concentration $U$ over time, as shown in Fig. \ref{stacked plot}(a). Based on Gaussian fits of the probability distributions obtained from random initialization, we observe that, at early times, two distinct peaks emerge ($t\,=100$), indicating the appearance of hybrid states. After a certain time, cells transitioned from their previous hybrid states to two additional extreme states (epithelial and mesenchymal), resulting quadrastability within the system ($t\,=500$). Eventually, the system stabilizes into another bimodal distribution corresponding to the two extreme epithelial and mesenchymal states at steady state ($t\,=2000$).\\
To further resolve the contribution of each phenotype, Fig. \ref{stacked plot}(b) shows the temporal evolution of the Gaussian distributions corresponding to epithelial, hybrid-I, hybrid-II, and mesenchymal states. Together, these results highlight the temporal progression of cells from hybrid phenotypes toward stable epithelial and mesenchymal states.

\subsection{Increased stability of Hybrid states:}
\begin{figure}[ht]
    \centering
    \includegraphics[width=1\linewidth]{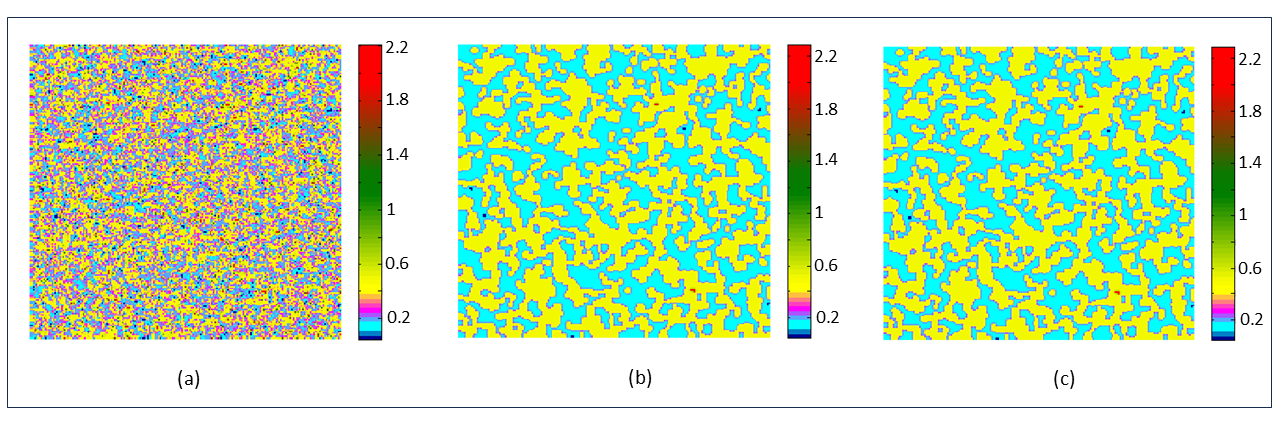}
    \caption{Increased stability of hybrid states in spatio-temporal dynamics. Corresponding parameters are: $K_{0u}=K_{0v}=0.005,\; K_{1u}=K_{1v}=0.4,\; K_u=K_v=0.09,\; d_{0u}=d_{0v}=0.9,\; K_c=1.8,\;K_{GR}=10.3, \;n=2,\; m=2$ Diffusion coefficients are $D_U=D_V=0.03$. Snapshots are taken after time (a) 10 (b) 1500 (c) 2000}
    \label{hybrid_stable}
\end{figure}

Recent experimental and theoretical studies have shown that hybrid epithelial/mesenchymal (E/M) phenotypes can be stably maintained, rather than being merely metastable, in both development and cancer contexts \cite{jolly2016stability,abell2011map3k4,lecharpentier2011detection}. \textcolor{black}{By stability, we refer to the dynamical persistence of the phenotype over time as an attractor of the underlying gene regulatory network.} These stabilized hybrid states are further associated with stemness, collective migration, and metastatic potential \cite{jolly2015coupling,hong2015ovol2}. Building upon these insights, we scan the parameter space further to explore possibilities of increased stability of Hybrid states. As it is known that hybrid cell peculiarities, such as collective migration, development and differentiation, regeneration, microenvironmental organization, metabolic adaptability, immunological evasion, immune suppression, and therapeutic resistance derive from their plasticity and stemness, we expect to find a parameter regime, where the cancer system will prefer the hybrid state, compared to extreme E or M states.   We observed the bifurcation of the protein $U$ with variation in the growth feedback rate [Fig. \ref{bif_dig} (d)] in quad-stable regime. Interestingly, in a higher rate of growth feedback, the intermediate hybrid states pose more stability than the extreme epithelial or mesenchymal cell types, as observable from Fig. \ref{bif_dig}(d). Spatiotemporally, under diffusive condition, hybrid I and hybrid II states result as final states, in spite of having possibilities of E or M state Fig. \ref{hybrid_stable}.

% \subsection{Time evolution \& Histograms:}
% To quantify the time evolution in terms of the spatial distribution, we have analyzed histogram plot illustrating the number of cells as a function of the concentration of protein $U$, as shown in Fig.\ref{histogram}. The analysis is performed with random initialization and diffusion coefficient set as $D_U$=$D_V$=0.04. Initially the system exhibited reaction driven dynamics and over time a bimodal distribution is observed indicating that cells bifurcate themselves into two intermediate states [Fig.\ref{histogram} (b)]. After a certain time cells transitioned from their previous hybrid states to two additional extreme states (epithelial and mesenchymal) resulting quadrastability within the system [Fig.\ref{histogram} (c)]. Eventually all cells from the previous intermediate state migrate towards the two extreme states leading to the formation of bimodal distribution of two extreme states [Fig.\ref{histogram} (d)].
% \begin{figure} [ht]
%     \centering
%     \includegraphics [width=1\linewidth]{Histogram_set3_final.png}
%     \caption{Histogram plot for number of cell-distribution with various concentrations of protein $U$. x-axis and y-axis represent the concentration of protein $U$ and the quantity of cells at that specific concentration respectively. Parameter values are taken as:$K_{0u}=K_{0v}=0.005, K_{1u}=K_{1v}=0.4, K_u=K_v=0.09, d_{0u}=d_{0v}=0.9, K_c=1.6,K_{GR}=9.25, n=m=2, D_U=D_V=0.04$.  Snapshots are taken after time: t=  (a)5  (b)100  (c)500  (d)2000 }
%     \label{histogram}
% \end{figure}
\subsection{Effect of Diffusion:}
\begin{figure}
    \centering
    \includegraphics[width=\linewidth]{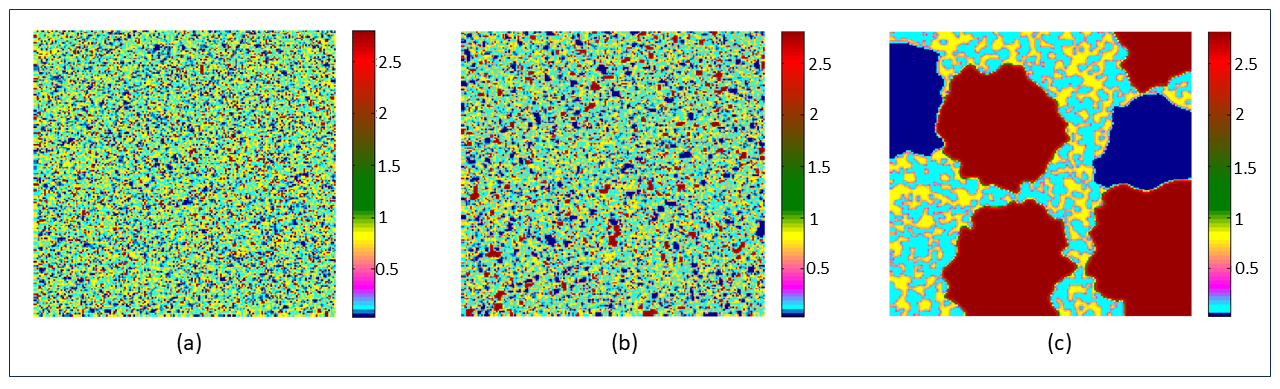}
    \caption{Spatiotemporal dynamics for different diffusion coefficient:(a) $D_U=D_V=0.005$ (b) $D_U=D_V=0.01$ (c) $D_U=D_V=0.05$ after time 1000. The other parameters are same as Fig. \ref{pattern}  }
    \label{differnt_diffusion}
\end{figure}

To investigate the effect of diffusion rate, we considered symmetric diffusion first; this implies, we consider both $U$ and $V$ have same diffusion rates, i.e., $D_U=D_V=D$. We tune $D$ to have low, moderate and high values to observe the effect on the 2D synthetic cell array. We show the results in Fig. \ref{differnt_diffusion}. We have two major observations here: first, the cluster/island sizes of each cell population are majorly driven by the diffusion rates. The greater the diffusion, the bigger the island sizes. Secondly, we note that hybrid states exist with more prominence for lower diffusion systems. Thus, the possibility of retaining the stemness and formation of small clusters is higher in lower diffusive systems.
\subsection{Asymmetric Diffusion \& spatiotemporal dynamics:}
\begin{figure} [ht]
    \centering
    \includegraphics [width=0.9\linewidth]{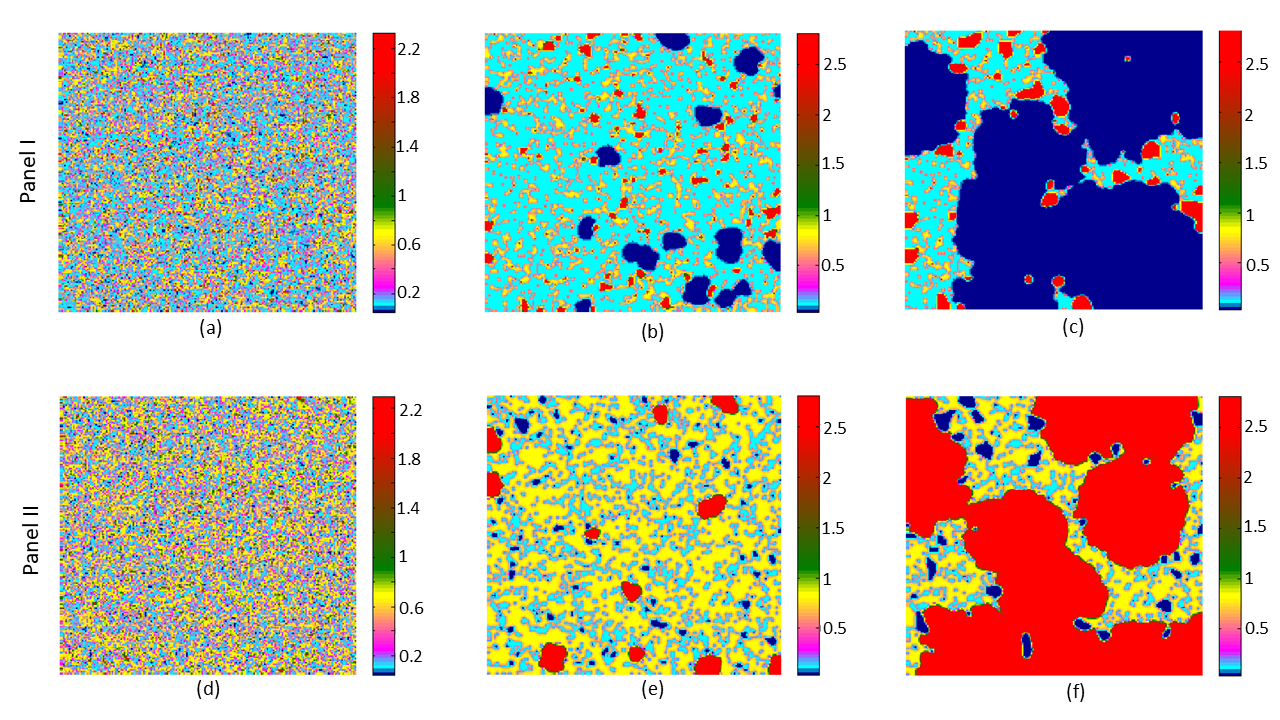}
    \caption{Diffusion controlled patterns for protein $U$. Panel I: Spatio-temporal pattern for $D_U=0.02$,\;$D_V=0.05$; snapshots are taken after time (a) 5 (b) 200 (c) 700; Panel II: Spatio-temporal pattern for $D_U=0.05,\;D_V=0.02$; snapshots are taken after time  (d) 5 (e) 200 (f) 700. Other parameters are same as Fig. \ref{pattern}.}
    \label{diffusion_driven_fig}
\end{figure}

%\textcolor{red}{ In the above discussion, we tuned the circuit parameters of Eq.  (3) and  (4) to observe how the system parameters spatially influence cell fate decisions.}In this section, we have illustrated that without altering gene expression parameters, how change in diffusion coefficients alone can lead the system to change the cell fate decision largely. 
In this section, we illustrate that even without altering gene expression parameters, changes in the diffusion coefficients alone can substantially shift cell fate decisions. For this, we simulated the system with random initialization  [$U_{initial}$  ($x_i$,$y_i$,0) = $\epsilon_1$\;$\xi$ (0,1),  $V_{initial}$  ($x_i$,$y_i$,0) = $\epsilon_2$\;$\xi$ (0,1)] considering the diffusion rate of protein $U$ and $V$ as different. In the Fig. \ref{diffusion_driven_fig} panel I picture set depicts the spatio-temporal pattern for lower $D_U$ and higher $D_V$  ($D_U=0.02,\;D_V=0.05$). Initially, the spatiotemporal dynamics is more reaction-driven [Fig. \ref{diffusion_driven_fig}(a)], after some time cells bifurcate into four different phenotypes with the hybrid-I and the lowest state (epithelial) being more prominent than the hybrid-II and highest state (mesenchymal) [Fig. \ref{diffusion_driven_fig}(b)]. As the system evolves with time the epithelial state becomes increasingly dominant as most of the cells switch themselves into this state [Fig. \ref{diffusion_driven_fig}(c)]. For panel-II, considering the diffusion rate for protein $U$ more than for protein $V$ ($D_U=0.05,\;D_V=0.02$), the spatiotemporal dynamics show the dominance of hybrid-II state and the highest state (mesenchymal) cells more than hybrid-I and the lowest state (epithelial) cells [Fig. \ref{diffusion_driven_fig}(e)]. Gradually, the system becomes more diffusion-driven and the mesenchymal state dominates the other states [Fig. \ref{diffusion_driven_fig}(f)]. This phenomenon suggests that the diffusion rate of different proteins influences the cells to switch their states among the four different phenotypes.  This phenotypic transition plays a crucial role in regulating the EMT progression and metastasis.           
\begin{figure}
    \centering
    \includegraphics[width=1.0\linewidth]{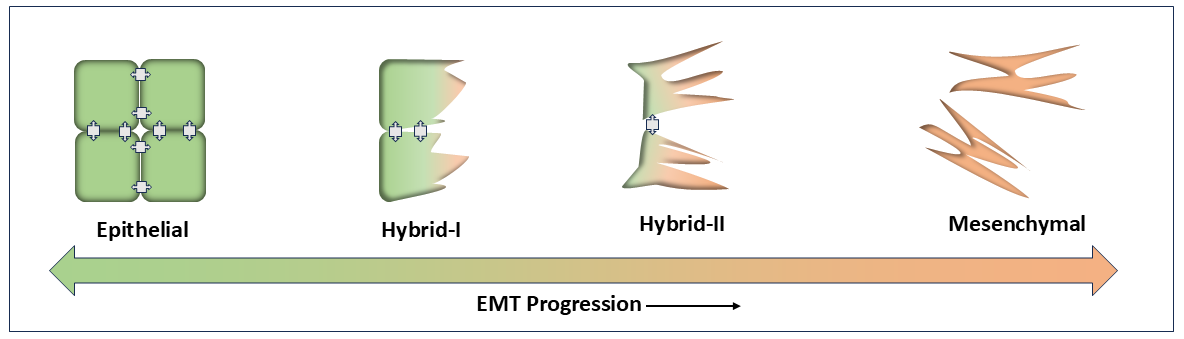}
    \caption{Schematic representation of EMT (left to right)/MET (right to left) progression. Cells transition from the epithelial state to the mesenchymal state through two intermediate hybrid phenotypes (hybrid-I and hybrid-II). The green-to-orange gradient arrow indicates the direction of EMT progression, with increasing loss of epithelial traits and gain of mesenchymal characteristics.}
    \label{EMT progression}
\end{figure}

\section{Discussions}\label{section 4}
In our study, we examine how emergent regulatory feedback, stemming from the growth burden of the
host, can lead to quad-stability within a simple bistable circuit. This study delves into the relatively unexplored realm concerning the impacts of spatial diffusion and pattern formation when there is a coupling between the host and the circuit. Within a tumor, where cells are densely packed, and resources
are limited, this growth burden poses a significant challenge. Interestingly, despite this crisis, the dynamics within the tumor are modulated in such a way that alternative states are created, exploiting nonlinearity. Consequently, cells within the tissue transition gradually between these states, which we
classify as hybrid-I and hybrid-II.
%In the Hybrid-I state, cells possess both adhesive properties and motility, albeit with a stronger adhesive tendency, anchoring them close to the primary tumor while allowing slight movement towards its periphery. Conversely, cells in the Hybrid-II state exhibit higher motility, forming smaller clusters within the tumor before eventually becoming fully motile and entering the bloodstream to establish secondary tumors elsewhere. These smaller clusters offer benefits such as increased drug resistance and enhanced mobility.\\
Our analysis suggests that resource competition plays a crucial role in controlling gene expression dynamics, particularly when growth feedback is present. When the resource competition is high due to cellular growth or cell division, a classical bistable genetic switch can exhibit multiple steady states. As growth plays a crucial role in the case of cancer \cite{witsch2010roles}, this multistable system generates the multiple phenotypes that dictate cancer dynamics and progression. It has been observed that tumor spheroid in a limited resource environment shows enhanced tumor growth, suggesting that resource competition intensifies the cancer progression and contributes to increased malignancy \cite{taylor2017resource,varahan2021correction}. In our analysis, the emergence of multiple hybrid states due to high resource competition further supports that idea. Moreover, our spatiotemporal observations indicate that the hybrid I and hybrid II states exhibit a higher propensity to transition toward purely epithelial or mesenchymal fates, whereas direct switching between epithelial and mesenchymal states is comparatively rare. This asymmetry in transition dynamics highlights the pivotal role of hybrid phenotypes as intermediates that facilitate EMT-like plasticity within the system. 
\\The theory that full EMT was linked to greater stemness has to be revised in light of the data that have been gathered thus far. In fact, it has now been mostly established that acquiring hybrid states improves stem-like characteristics. A delicate balance governs the regulation of hybrid states, enabling both the stabilization of hybrid intermediates and the advancement back and forth towards an entirely E or M state \cite{jolly2016stability,canciello2022medio,tedja2023generation}. The loss of the E phenotype is accompanied by a rise in stemness and plasticity, which peaks during the creation of hybrid states and then declines once EMT is finished. Fig. \ref{EMT progression} schematically represents EMT progression, showing four cell types transitioning between states, and the transition among these phenotypes is crucial for cell dissemination and colonisation at distant tissues to enhance the metastatic potential. 
\textcolor{black}{
\section{Conclusions}
In summary, our findings demonstrate that even a minimal genetic circuit, when coupled to host growth dynamics and resource competition, can give rise to quad-stability and multiple hybrid phenotypes. These results highlight how emergent feedback from resource limitation alone—without invoking complex genetic regulatory networks—can promote phenotypic heterogeneity and stabilize intermediate EMT states. Such hybrid states play a central role in enabling cellular plasticity and mediating transitions between epithelial and mesenchymal phenotypes, contributing to latent tumors and posing a substantial hurdle in treatment.}
 Our discoveries hold immense importance in understanding the drug resistance mechanisms of cancer cells.  Developing drugs that specifically target and limit this intermediate state, there's a promising avenue for treating cancer effectively. Targeted therapies aimed at disrupting the transition between cancer cell states could potentially arrest the advancement of the disease, offering hope for improved cancer treatments.\\
 \textcolor{black}{ Incorporation of the distinct mechanical, migratory and adhesive properties that define epithelial ( stable, polarized, collective organization) and mesenchymal (motile, plastic, extracellular matrix-interactive) phenotypes represents a natural extension of the current framework. Future studies integrating cellular adhesion, cytoskeletal dynamics, and extracellular matrix interactions will be essential to bridge these regulatory states with their corresponding physiological behaviors. Such extensions would enable a more comprehensive mapping between regulatory states and functional phenotypes during cancer progression.}
 %the physiological traits of Epithelial, hybrid and mesenchymal cells (like, stable, polarized, collective organization vs. motility, plasticity, and migration) that address mechanical, migratory, and extracellular matrix-dependent properties can be addressed in a future work integrating cellular adhesion and movement.
 
%\section*{Data availability statement}
%The datasets analyzed during the current study are available in the KONECT repository, ttp://konect.cc/networks/arenas-jazz/.
\section*{Conflict of Interest}The authors declare that they do not have any known conflicts of interest.
\section*{Data availability} No dataset is used in conducting the research described in this article.
\section*{Acknowledgements}
RK acknowledges the support by DST-INSPIRE, India, vide sanction Letter No. DST/INSPIRE Fellowship/2022/IF220269  dated- 22.03.2024.
%%=============================================%%
%% For presentation purpose, we have included  %%
%% \bigskip command. Please ignore this.       %%
%%=============================================%%

%%=============================================%%
%% For presentation purpose, we have included  %%
%% \bigskip command. Please ignore this.       %%
%%=============================================%%

%%=============================================%%
%% For presentation purpose, we have included  %%
%% \bigskip command. Please ignore this.       %%
%%=============================================%%

%%=============================================%%
%% For presentation purpose, we have included  %%
%% \bigskip command. Please ignore this.       %%
%%=============================================%%


\bibliography{sn-bibliography}% common bib file

\end{document}